\def\etal{{\it et al.}\/}

\documentclass[aps,prb,showpacs,superscriptaddress,a4paper,twocolumn]{revtex4-1}

\usepackage{amsmath}
\usepackage{color}
\usepackage{float}
\usepackage{amssymb}
\usepackage{graphicx}
\usepackage{graphics}
\usepackage{xcolor}
\usepackage{blindtext}

\begin {document}

\title{Revealing the nature of antiferro-quadrupolar ordering in Cerium Hexaboride: CeB$_6$}
\author{C.K. Barman}  \email{chanchal$_$barman@iitb.ac.in}
\affiliation{Department of Physics, Indian Institute of Technology, Bombay, Powai, Mumbai 400 076, India}
\author{Prashant Singh}   \email{corresponding author: prashant40179@gmail.com}
\affiliation{Ames Laboratory, U.S. Department of Energy, Ames, Iowa 50011, USA}
\author{Duane D. Johnson}  \email{ddj@iastate.edu}
\affiliation{Ames Laboratory, U.S. Department of Energy, Ames, Iowa 50011, USA}
\affiliation{Materials Science $\&$ Engineering, Iowa State University, Ames, Iowa 50011, USA}
\author{Aftab Alam} \email{aftab@phy.iitb.ac.in}
\affiliation{Department of Physics, Indian Institute of Technology, Bombay, Powai, Mumbai 400 076, India}

\begin{abstract}
Cerium-hexaboride (CeB$_6$) f-electron compound displays a rich array of low-temperature magnetic phenomena, including `magnetically hidden' order, identified as multipolar in origin via advanced x-ray scattering. From first-principles electronic-structure results, we find that the \emph{antiferro-quadrupolar} (AFQ) ordering in CeB$_{6}$  arises from crystal-field splitting and yields band structure in agreement with experiments. With interactions of $p$-electrons between Ce and B$_{6}$ being small, the electronic state of CeB$_{6}$ is suitably described as  Ce(4$f^{1}$)$^{3+}$(e$^{-}$)(B$_{6}$)$^{2-}$.
The AFQ state of orbital spins is caused by an exchange interaction induced through spin-orbit interaction, which also splits J=5/2 state into $\Gamma_{8}$ ground state and $\Gamma_{7}$ excited state. Within the  smallest antiferromagnetic (111) configuration, an orbital-ordered AFQ state appears during charge self-consistency, and supports the appearance of `hidden' order. Hydrostatic pressure (either applied or chemically induced) stabilizes the AFM (AFQ) states over a ferromagnetic one, as observed at low temperatures.
\end{abstract} 
\date{\today}
\maketitle

{\par}The nature and first-principles description of $f$-electron materials is a considerable challenge and a highly debated topic in condensed-matter physics. 
The simultaneous presence of itinerant $s$-$p$-$d$ states and partially occupied localized $f$-states and their interaction in rare-earth materials give rise to a rich variety of phenomena, and remain a serious test for electronic-structure theories.\cite{Casadei2012}  Rare-earth compounds with 4$f$ electrons possessing orbital plus spin degrees of freedom generally show electric quadrupole ordering in addition to magnetic dipole ordering at low temperatures.\cite{1,2}  In cerium-based compounds, the single 4$f$ electron gives rise to anomalous and fascinating behavior, such as heavy-fermion, intermediate valence compounds, Kondo metals, and Kondo insulators.\cite{a,b,c,d,e}

{\par}Cerium hexaboride (CeB$_{6}$) is considered as a typical example of an $f$-electron system, where Ce$^{+3}$ ions are arranged in the simple cubic lattice and quadrupolar interactions play an important role in its magnetic behavior.\cite{3,4} 
It shows a unique antiferro-quadrupolar (AFQ) ordering\cite{PS2009,b} at temperature T${_Q}< 3.2$ K, associated with ordering of magnetic quadrupolar moments at cube corners.\cite{MSHI2001,HJ2014} 
Quadrupolar ordering is orbital in nature, arising due to the distortion of electronic charge density of the unpaired electrons in their 4$f$ orbitals. 
The AFQ ordering has also been observed in compounds like DyB$_{2}$C$_{2}$, HoB$_{2}$C$_{2}$, TmTe and PrOs$_{4}$Sb$_{12}$.\cite{2,5,6,7,8}

{\par}The ordering phenomena in CeB$_6$ is acknowledged to be governed by antiferromagnetic (AFM)\cite{AS2013,OZ2003} interactions between multipolar moments of  Ce-4$f$ electrons mediated by itinerant conduction electrons, which lift the degeneracy of the $\Gamma_8$ state of the Ce ions in their cubic crystal field.\cite{a,PT2003}
Although the energy of the 4$f$ electron is in the range of 5$d$ and 6$s$ valence electrons, its wave function is spatially localized and tighter than semicore 5$s$ and 5$p$ electrons. 
The competition between Ce 4$f$ electron being itinerant or localized determines the character of the compounds. 
The challenge is to describe coexistent near-degenerate, low-temperature  phases\cite{JME1985,NS1984,TSSS1998, SSS1999} that arise from Ce-4$f$ hybridization with B conduction electrons. 
Jang, \emph{et al.}\cite{HJ2014} highlighted the ferromagnetic (FM) correlations in CeB$_{6}$ and suggested an intimate interplay between orbital\cite{HN2001,TM2009,TM2012} and magnetic ordering.\cite{AS2013,OZ2003} 


{\par} 
To investigate AFM (magnetic) and AFQ (charge) ordering, we explore the electronic structure using first-principles density functional theory (DFT) with increased orbital (charge) and magnetic degrees of freedom, and find close agreement with experiments.\cite{28} Using a 2$\times$2$\times$2 supercell with inequivalent Ce atoms, we use DFT as implemented in the Vienna Ab-Initio Simulation Package (VASP)\cite{KH1993,KJ1999}  to permit different charge and magnetic periodicities.
The valence interactions were described by projector augmented-wave method \cite{PEB1994,KJ1999} with energy cutoff of 320 eV for the plane-wave orbitals. 
We use 7$\times$7$\times$7 Monkhorst-Pack $k$-mesh for Brillouin zone sampling.\cite{MP1976} 
Total energies were converged to $10^{-5}$ eV/cell. 
We employ the Perdew-Bueke-Ernzerhof (PBE)\cite{PBE1997} exchange-correlation functional in the generalized gradient approximation (GGA). 
In (semi)local functionals, such as GGA, the $f$-electrons are always delocalized due to their large self-interaction error. 
To enforce localization of the $f$-electrons, we perform PBE+U calculations\cite{PBE+U} with a Hubbard U (3 eV; J=1 eV) introduced in a screened Hartree-Fock manner.\cite{hubbardU} 
The relativistic spin-orbit coupling (SOC) is also included and provides an interaction between the atomic orbital angular momentum and electron spin, a small perturbation of electrons in solids except for heavy elements with $f$-orbitals, where it need not be weak -- it effectively increases proportionally to $Z^{4}$ ($Z$ is atomic number).  
 
\begin{figure}[t!]
\centering
{\includegraphics[scale=0.38]{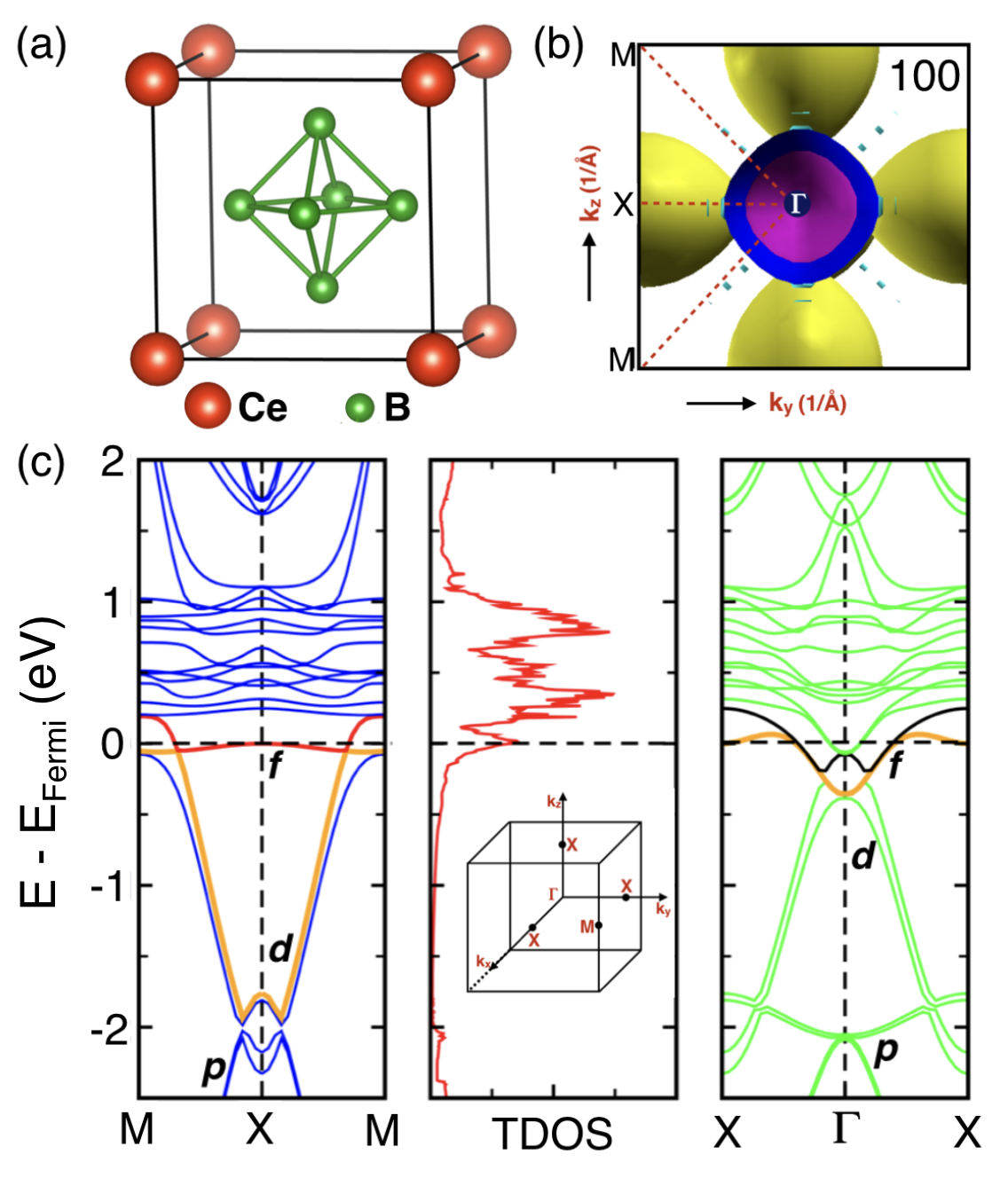}}
\caption {For FM CeB$_{6}$, (a) $Pm{\bar{3}}m$ crystal structure, (b) (100) Fermi-surface, and (c) bands (with SOC) along M-X-M and X-$\Gamma$-X, with $p$, $d$, and $f$ states identified, and density of states (DOS). Inset: Brillouin zone and high-symmetry points.}\label{Fig1}
\end{figure}

{\par}{\it Crystal structure \& valency:}~CeB$_{6}$ possesses a unique simple-cubic structure (space group  $Pm\bar{3}m$) comprised of Ce$^{3+}$ ions separated by B$_{6}$ octahedra, see Fig.~\ref{Fig1}(a), with the lattice constant $a$ of 4.14~$\AA$.\cite{LL2011} The calculated lattice constant of 4.147~$\AA$ shows good agreement with experiments.\cite{LL2011} 
The structure can be considered as two interpenetrating simpl-cubic sublattices, one consisting of B$_{6}$ octahedra and the other of Ce ions. 
Notably,  B$_{6}$ octahedra, which form a covalently bonded structure, require two additional electrons from the Ce ions to be stabilized.\cite{YSG1985,HCL1954} 
The Ce valence is $\left[Xe\right]$4${{f}^{1}}$5${{d}^{1}}$6${{s}^{2}}$, with the two $s$-electrons being donated to the B$_{6}$ octahedra, and it is generally considered that the $f$-electron states remain localized with the $d$-electrons forming the conduction band, resulting in the Ce$^{3+}$ ion. 
Grushko \etal~compared the X-ray chemical shift with the self-consistent Dirac-Fock-Slater-Latter calculation and concluded that the trivalent rare-earth atoms in hexaboride with metallic conduction donate two electrons to the boron framework, and that a third valence electron exists in the 5d orbitals. \cite{YSG1985}

{\par}Initially Ce$^{3+}$ multiplet 4$f^{1}$ was thought to be split by the crystalline electric field into a ${{\Gamma}_{7}}$  ground state with a ${{\Gamma}_{8}}$  excited state.\cite{JCN1969} 
However, this was later reversed to the ${{\Gamma}_{8}}$ quartet ground state, which is four-fold degenerate with 2-orbital and 2-spin degrees of freedom, located 46 meV below the $\Gamma_{7}$ doublet state.\cite{EZ1984, ML1985, NS1984} 
Our present results give $\sim$62~meV separation between $\Gamma_{7}$ and $\Gamma_{8}$, reflecting experimentally findings. 
The Raman scattering measurements provide an explanation for these observations, indicating that the ${{\Gamma}_{8}}$  quartet is further split into two doublets, ${{\Gamma}_{8,1}}$  and ${{\Gamma}_{8,2}}$ , separated by around 30 K.\cite{EZ1984}

{\par} In Fig.~\ref{Fig1}(c), we show the band structure and  density of states (DOS) for FM states. 
The low-T phase and its electronic structure is mainly governed by the dispersive 5$d$ and flat 4$f$ bands, shown along M-X-M and X-$\Gamma$-X. 
The flat bands near the Fermi-energy (E$_{Fermi}$) arise purely from Ce-4$f$ states. 
The dispersive $d$-band (X points) is found to be about ${-2.0}$ eV below E$_{Fermi}$ and the dispersive B $2p$ bands are near the bottom of this $d$-band. 
These bands at or near the X-point agree fairly well with experiments.\cite{28,AK2016} 
One  immediately notices the location of flat Ce-4$f$ bands slightly below E$_{Fermi}$, as observed in ARPES data,\cite{28} although their energy position differ slightly. 
DOS shows similar behavior, but with most significant density below E$_{Fermi}$.
Importantly, a parabolic band along X-$\Gamma$-X  forms close to E$_{Fermi}$ at $\Gamma$ giving a hole-like pocket, as observed.\cite{28,AK2016}
A strong renormalization of bands near E$_{Fermi}$ at $\Gamma$-point occurs in both these cases. 
Several features in these bands  can be corroborated with the ARPES data.\cite{28,AK2016} 
Parabolic shaped bands near E$_{Fermi}$ at $\Gamma$-point which are relatively more flat  compared to those in ARPES data.\cite{28,AK2016} 
In contrast to previous calculations,\cite{28,AK2016} we find hole-like character near  at $\Gamma$. 
The calculated Fermi-surface, Fig.~\ref{Fig1}(b), is in good agreement with observation,\cite{28,AK2016} i.e., hole pockets, including an oval-shaped contour at X, are found. 
The spectral intensities around $\Gamma$ are stronger compared to those at X. 
The two Fermi-surface contours (blue and magenta around $\Gamma$ in Fig.~ \ref{Fig1}(b)) represents the band splitting. 
In Fig.~\ref{Fig1}(b), the hole-like pocket at $\Gamma$, with strongly renormalized bands, correspond the observed, so-called, \emph{hot spots}.\cite{28,AK2016}
The emergence of low-temperature magnetic order is highly possible if these states are extremely close in energy relative to the FM case.

\begin{figure*}[t!]
\centering
\includegraphics[scale=0.4]{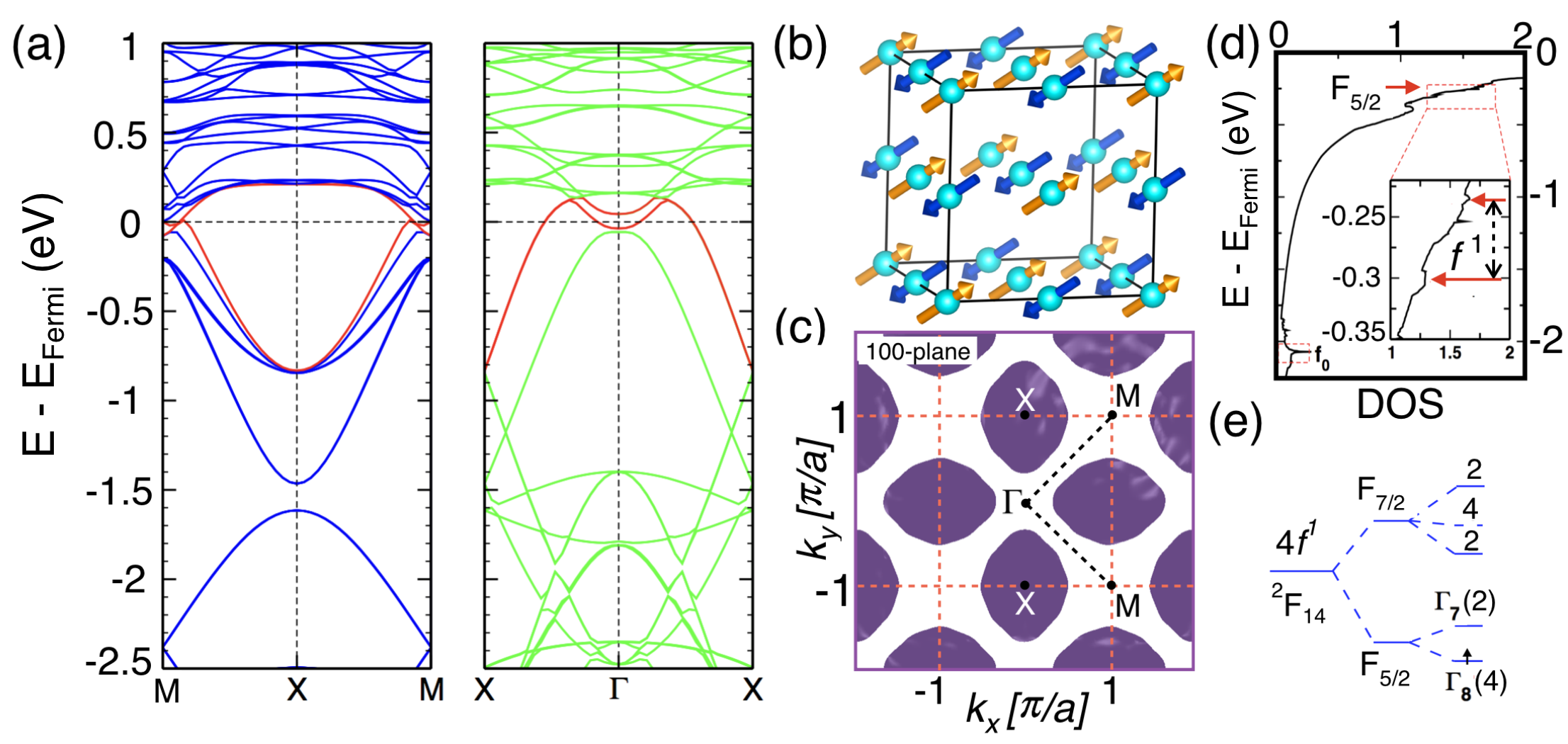}
\caption {For AFM CeB$_{6}$, (a) DFT+U dispersion along M-X-M and X-$\Gamma$-X. (b) Schematic of AFM state. (c) (100) constant 
energy surface plot at -0.30~eV below Fermi-energy ($\Gamma_{7}$ and $\Gamma_{8}$ splitting is observed in -0.20 to -0.35 eV energy range). (d) Ce$-f$ DOS (matches experiment).\cite{28} (e) Energy-level diagram with SOC and crystal-field splitting.}\label{Fig2}
\end{figure*}

CeB$_{6}$ has been investigated intensively at low temperature due to its unusual properties such as AFQ ordering, the Kondo effect (which makes the localized Ce moment vanish due to  coupling of Ce and B moments), and the Ruderman-K\"uttel-Kasuya-Yoshida (RKKY) interaction (which arranges the moments of the AFM Ce with the moments of itinerant electrons of B).\cite{sato85, sakai97} 
These properties are closely connected with the localized 4$f (\Gamma_{8}$) electrons of Ce and conduction electrons of B. 
Including SOC interactions resolves most of the differences except the presence of AFQ-type charge ordering. 
To elucidate on AFQ phase, we perform similar SOC+U calculation on larger supercell with AFM arrangement of spins on Ce1 [along (111)] and Ce2 [along -(111)] see Fig.~\ref{Fig2}(b). 

{\par}From the band structure of CeB$_{6}$, Fig.~\ref{Fig2}(a), the 4$f$ bands are hybridized with the 5$d$ band around E$_{Fermi}$  \cite{heide86} of CeB$_{6}$. 
The 4$f$ bands are centered at the X point in the Brillouin Zone, which is the center of the B$-$B bond and has hybridized character of Ce-5$d$ and B-2$p$ states.\cite{souma04,4_a}  
The calculated constant energy surface plot in Fig.~\ref{Fig2}(c) at -0.30 eV below E$_{Fermi}$ consists of ellipsoids centered about the X point, and are typical for the hexaborides,\cite{28,AK2016} in agreement with previous measurements.\cite{souma04,28,AK2016,30}  
Their ellipsoid orbital character is composed of extended Ce-5$d$ states with admixtures of localized Ce-4$f$ near E$_{Fermi}$, similar to other 4$f$ systems exhibiting a resonance mode.\cite{31,32} 
The large electron-like constant energy surface plot centered at X (M) point is in good agreement with experiments.\cite{souma04,4_a} 
The ellipsoidal-shape in constant energy surface plot, elongated along the X(M) - $\Gamma$ (X), does support the assumptions of the two models used to explain the AFQ and AFM ordering in CeB$_{6}$.\cite{Kurmamoto02,PS2012}

{\par}The valence-band structure along M$-$X$-$M and X$-\Gamma-$X direction is shown in Fig.~\ref{Fig2}(a) in the cubic Brillouin zone.\cite{4_a} 
We find that the gross feature of band structure is in good agreement with existing experiments. 
According to the band calculation, the observed dispersive bands in this energy range are attributed to the bonding B 2$s-$2$p$ state of the  octahedron. 
Also, the non-dispersive band at 2.1 eV belongs to Ce-$d$ states. 
The band along X$-\Gamma-$X direction has a parabolic (or U) shape, whereas the bottom of the band appears more cusp-like (or V) shape along M$-$X$-$M. 
Near E$_{Fermi}$, the screened $f^{1}$ states are found, which split due to the spin-orbit coupling in a J equals 5/2 and 7/2 component. 
The 5/2 state at E$_{Fermi}$ is relevant here and splits further into crystal-field levels under SOC and DFT+U, see Fig.~\ref{Fig2}, namely, a $\Gamma_{7}$ doublet (excited state) and a $\Gamma_{8}$ quartet.\cite{246} 
One of the $\Gamma_{8}$  levels is occupied, whereas the $\Gamma_{7}$  intensity seen in the spectrum is a satellite. 
The energy separation of the $\Gamma_{7}$  and $\Gamma_{8}$  levels ($62~me$V) is in agreement with previous reports.\cite{33,34} 
Note that the large ground-state degeneracy distinguishes CeB$_{6}$ from many other Ce-based heavy-fermion materials.

{\par}The 4$f$ state in Ce ions with stable valency has the orbital freedom in addition to the spin. 
The ground state multiplet due to the spin-orbit interaction splits into the crystalline electric field state by the multipolar Coulomb potential. 
As shown in the level splitting, in the $f^{1}$ configuration, $\Gamma_{8}$ is lower than $\Gamma_{7}$. 
In Fig.~\ref{Fig2}(d), the localized $f^{0}$ ionization peak of Ce-$f$ at $-2.05~e$V overlaps with the bottom of the ellipsoid band, and agrees  with those of the integrated energy distribution from experiments.\cite{28,AK2016} 
Below E$_{Fermi}$, the screened $f^1$ states of Ce, located between $-$0.2 to $-$0.35 eV, splits into J $5/2$ and $7/2$ components due to SOC. 
Interestingly, the 4$f$($j$ = 5/2) orbital further splits into $\Gamma_{7}$ (doublet) excited states and $\Gamma_{8}$ (quartet) ground states under $O_{h}$ crystal field. 
To emphasize, for $\Gamma_{8}$ to be ground state, the SOC interaction should be larger than the Hund's rule interaction.\cite{253,254}  
As such, the energy level of the 4$f$(5/2) orbitals remains lower than the 4$f$(7/2) orbitals. Ce$^{3+}$ formally has one 4$f$-electron. 
The $\Gamma_7$ and $\Gamma_8$ differ in energy by $62~me$V, agreeing fairly well with the $50~$meV from photoemission. \cite{28,AK2016} 

{\par}The schematic energy levels are illustrated in Fig.~\ref{Fig2}(e).\cite{M2008} In spite of same local crystal-field anisotropy in AFM CeB$_{6}$, the opposite moments on Ce1 and Ce2 results in no gain in energy due to the magnetic dipole interaction. This unusual magnetic structure is now understood to be a consequence of the underlying AFQ order, which confines the direction of the magnetic moment by a strong spin-orbit coupling. 

\begin{figure}[t]
\centering
\includegraphics[scale=0.3]{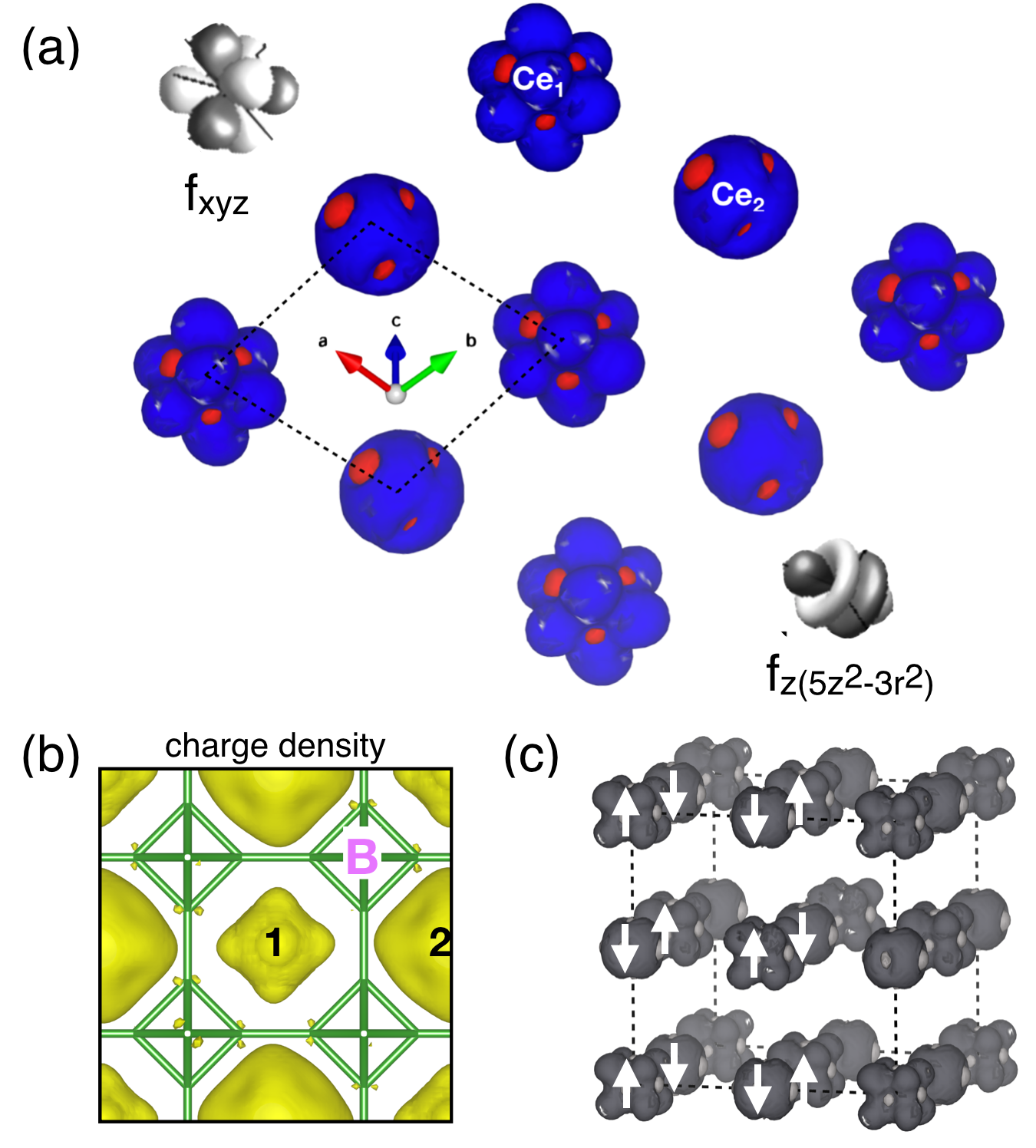}
\caption {For AFM CeB$_{6}$, we show (a) (111)-projected Ce1 ($f_{xyz}$) and Ce2 ($f_{z(5z^{2}-3r^{2})}$) orbitals, (b) total charge density in (001) plane, and (c) schematic of AFM Ce configuration. Together these show distinct AFM and orbital arrangement at Ce1 and Ce2 sites, indicating the underlying AFQ order.}\label{Fig3}
\end{figure}

In AFQ CeB$_{6}$, the 4$f$-electrons are localized, having an orbitally degenerate level in the crystalline electric-field ground state. 
As shown in Fig.~\ref{Fig3}, the orbital ordering in $f$-electron systems, i.e., a spontaneous lifting of the orbital degeneracy, is a phase transition of quadrupole moments. The orbital degeneracy then is described in terms of quadrupole moments due to presence of strong intra-atomic SOC. 
Following the AFM ordering, one refers to uniform alignment of the quadrupole moment, where this staggered quadrupolar component is called an AFQ state. The effect is also visible in Fig.~\ref{Fig3}(b) through contrasting charge density at Ce1 and Ce2 sites. For CeB$_{6}$, an AFM state with an AFQ background is evident in Fig.~\ref{Fig3}(c).

From the axial interaction with B-$p$ states, the Ce-$f$ states (f$_{xyz}$ and f$_{z(5z^{2}-3r^{2})}$) are modified and produce a weak electric quadrupolar ordering with (nearly) degenerate localized states. The charge distributions on Ce1 and Ce2, in Fig.~\ref{Fig3}(b), comes from f$_{xyz}$ and f$_{z(5z^{2}-3r^{2})}$ orbitals, respectively, giving distinct shape to the charge density. This underlying (``hidden'') AFQ ordering is difficult to observe as this arises mainly from weaker quadrupolar interaction and the electron density in the given unit cell spontaneously distort in a repeating pattern throughout the crystal.

\begin{figure}[t]
\centering
\includegraphics[scale=0.35]{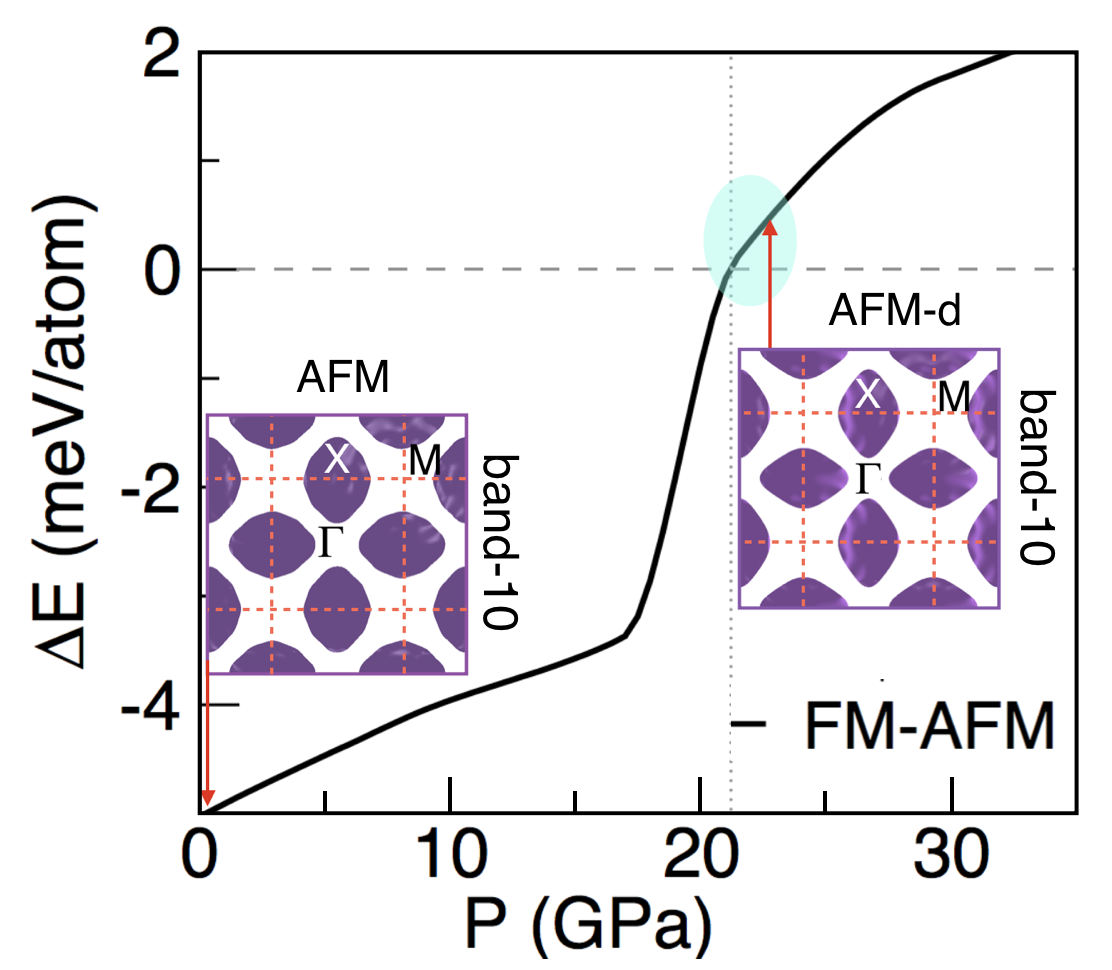}
\caption {For CeB$_{6}$, FM$-$AFM energy difference vs. pressure, which alters hybridization between 4$f$-bands and dispersive 5$d$-bands (Inset: AFM constant energy surface (hole-states) at -0.30 eV below E$_{Fermi}$ at 0 and 21 GPa. Hole-like states appear at $\Gamma$ and X points (see Fig. S8)). \cite{28}}\label{Fig4}
\end{figure}

For any admixture of magnetic-dipole, charge-order or sufficiently large lattice distortion, the neutron scattering shows indirect coupling to the multipolar order but remains unchanged in quadrupolar AFQ phase.\cite{5_1} 
URu$_{2}$Si$_{2}$ is one such example.\cite{8_1} 
In Fig.~\ref{Fig4}, we show the effect of (hydrostatic) pressure on the relative energy of FM and AFM states, where they are degenerate near 21 GPa (-2.5\% change in lattice constant), above which the AFM is stable. The simulated energy difference between FM and AFM phase lie within few meV (1 meV is equivalent to 11 Kelvin). Such small energy difference sometimes acts as the precursor for magnetic phase instability and infers the co-existence of magnetic domains. This point is carefully taken up in a recent study using high intensity inelastic neutron scattering [Ref. 13].~The competition between FM and AFM states is sensitive to  pressure due to the hybridization between flat 4$f$-bands and low-lying dispersive 5$d$-bands, as reflected in the constant energy surface plot changes in shape and size of the hole-pockets at the X-point (Fig.~\ref{Fig4}). However, the pattern is similar to those observed by Neupane {\it et al.},\cite{28} and clearly shows the presence of hole-like states (X-point). 

In summary, we have provided direct electronic insight to the presence of antiferro-quadrupolar (AFQ) ordering in CeB$_{6}$.
The crystal-field splitting, controlled by spin-orbit coupling, yield electronic dispersion and constant energy surface below E$_{Fermi}$ (electron and hole pockets) that agree fairly well with those observed from ARPES, highlighting the importance of spin-orbit coupling in $f$-block systems. 
Furthermore, our calculations reveal that dispersion around  $\Gamma$ is strongly renormalized, as indicated by highly increased density of states there, which are observed as hot-spots in ARPES. The competition between FM and AFM states is sensitive to pressure (both applied and chemically induced), which alters the hybridization between flat 4$f$-bands  and low-lying dispersive 5$d$-bands.
Finally, with a recent finding of topologically insulated phase in SmB$_{6}$, a  search for topological insulator phase with magnetically active sites in CeB$_{6}$ may be  warranted.

\section*{Acknowledgement}
CKB and PS equally contributed to this work. 
We thank Rebecca Flint and Peter Orth at Iowa State University for clarifying discussions.
CKB was supported from teaching assistantship at IIT Bombay. 
PS and DDJ were funded by the U.S. Department of Energy (DOE),  Office of Science, Basic Energy Sciences, 
Materials Science and Engineering Division. Ames Laboratory is operated for the U.S. DOE by Iowa State University 
under Contract No. DE-AC02-07CH11358.


\end{document}